\documentclass[twocolumn,superscriptaddress,amsmath,amssymb,showpacs,prb]{revtex4-2}

\usepackage{graphicx}
\usepackage{color}
\renewcommand{\phi}{\varphi}

\begin{document}

\title{Iron-rich Fe-O compounds with closest-packed layers at core pressures}

\author{Jin Liu}
	\thanks{These authors contributed equally.}
	\affiliation{Center for High Pressure Science and Technology Advanced Research, Beijing 100094, China}
	\affiliation{CAS Center for Excellence in Deep Earth Science, Guangzhou, 510640, China}
\author{Yang Sun}
	\thanks{These authors contributed equally.}
	\affiliation{Department of Applied Physics and Applied Mathematics, Columbia University, New York, NY 10027, U.S.A.}
\author{Chaojia Lv}
		\affiliation{Center for High Pressure Science and Technology Advanced Research, Beijing 100094, China}
\author{Feng Zhang}
	\affiliation{Department of Physics, Iowa State University, Ames, Iowa 50011, U.S.A.}
\author{Suyu Fu}
	\affiliation{Department of Geological Sciences, Jackson School of Geosciences, The University of Texas at Austin, Austin, TX 78712, U.S.A.}
\author{Vitali B. Prakapenka}
	\affiliation{Center for Advanced Radiation Sources, University of Chicago, Chicago, IL 60439, U.S.A.}
\author{Cai-Zhuang Wang}
	\affiliation{Department of Physics, Iowa State University, Ames, Iowa 50011, U.S.A.}
\author{Kai-Ming Ho}
	\affiliation{Department of Physics, Iowa State University, Ames, Iowa 50011, U.S.A.}
\author{Jung-Fu Lin}
	\affiliation{Department of Geological Sciences, Jackson School of Geosciences, The University of Texas at Austin, Austin, TX 78712, U.S.A.}
\author{Renata M. Wentzcovitch}
    \email [Email: ]{jinliuyc@foxmail.com (J.L.)\\afu@jsg.utexas.edu (J.F.L.)\\ rmw2150@columbia.edu (R.M.W.)}
	\affiliation{Department of Applied Physics and Applied Mathematics, Columbia University, New York, NY 10027, U.S.A.}
	\affiliation{Department of Earth and Environmental Sciences, Columbia University, New York, NY 10027, U.S.A.}
	\affiliation{Lamont–Doherty Earth Observatory, Columbia University, Palisades, NY 10964, U.S.A.}

\begin{abstract}

Oxygen solubility in solid iron is extremely low, even at high pressures and temperatures. Thus far, no Fe-O compounds between Fe and FeO endmembers have been reported experimentally. We observed chemical reactions of Fe with FeO or Fe$_2$O$_3$ \textit{in situ} x-ray diffraction experiments at 220-260 GPa and 3,000-3,500 K. The refined diffraction patterns are consistent with a series of Fe$_n$O (n $>$ 1) compounds (e.g., Fe$_{25}$O$_{13}$ and Fe$_{28}$O$_{14}$) identified using the adaptive genetic algorithm. Like $\epsilon$-Fe in the hexagonal close-packed (hcp) structure, the structures of Fe$_n$O compounds consist of oxygen-only close-packed monolayers distributed between iron-only layers. \textit{Ab initio} calculations show systematic electronic properties of these compounds that have ramifications for the physical properties of Earth's inner core.

\end{abstract}

\date{Oct. 1, 2021}

\maketitle

Iron is the most abundant transition metal in the solar system and the primary constituent of Earth's metallic core, while oxygen is the most abundant element in the solid Earth \cite{1,2}. Fe and O together account for $>$ 60$\%$ of the Earth's total mass and play dominant roles in the planet's evolution, including the origin of the geodynamo. The mixed valence metal Fe allows for several naturally occurring iron oxides, FeO, Fe$_3$O$_4$, and Fe$_2$O$_3$. Notably, wüstite (Fe$_{1-x}$O, 0 $<$ x $<$ 0.15) is non-stoichiometric because of the presence of both Fe$^{2+}$ and Fe$^{3+}$ states \cite{3}. On the other hand, oxygen, a lithophile element in column VI, has very low solubility in solid iron even at high pressures and temperatures (P-T) and can remain highly immiscible in liquid iron \cite{4,5}. Thus far, the occurrence of solid Fe$_n$O (n $>$ 1) stoichiometric compounds remains an open question in the Fe-FeO system.

At high P-T conditions, new compounds between two insoluble elements are well known to form with drastically different physical properties \cite{6,7}. For example, sulfur, another element in group VI, forms two iron-rich metallic sulfides, Fe$_2$S and Fe$_3$S, at tens of gigapascals (GPa) and high temperatures \cite{8,9}. Several oxygen-rich Fe-O compounds have been synthesized at high P-T in the Fe-O binary system, including Fe$_4$O$_5$, Fe$_5$O$_6$, Fe$_5$O$_7$, Fe$_2$O$_{3+\delta}$, FeO$_2$, and Fe$_{25}$O$_{32}$ \cite{10,11,12,13,14}. Previous \textit{ab initio} calculations on Fe$_3$O and Fe$_4$O, an iron-rich Fe-O compound, did not find it to exhibit a stable structure at pressures up to 350 GPa at 0 K \cite{9,15}. In contrast, recent crystal structure searches predicted low-enthalpy Fe$_3$O phases at 350 and 500 GPa, respectively \cite{16}. On the other hand, experimental studies have not shown any stable Fe$_n$O phases at pressures up to 100-200 GPa \cite{5}. At extreme conditions, previous studies have found that FeO becomes metallic at $\sim$ 70 GPa and 1900 K, featuring close-packed structures with alternating iron-only and oxygen-only monolayers \cite{17}. Pure oxygen also becomes metallic at 96 GPa and room temperature \cite{18}. These observations support the notion that the chemical characters of Fe and O change under pressure, possibly changing oxygen solubility in solid iron \cite{4}. Besides, oxygen is a likely light element alloyed with Fe in Earth's core. As much as $\sim$10 wt.$\%$ light element(s) in liquid iron could partition between the liquid outer and solid inner core at approximately 330 GPa and 5500 K \cite{5,19}. The consensus from previous studies is that oxygen remains a lithophile element similar to Mg and Al with very limited solubility in solid iron, instead of becoming a siderophile element like S, even at extremely high P-T conditions of the core \cite{20}. As such, most oxygen would remain in the liquid during inner core growth, providing gravitational energy and latent heat as key energy sources powering the geodynamo. Consequently, oxygen has not been proposed yet to be present as Fe$_n$O in the solid inner core. Knowledge on the stability and physical properties of Fe$_n$O compounds at relevant core P-T conditions is thus of great significance to our understanding of the geophysics of the planet.

We combined high P-T x-ray diffraction (XRD) experiments and crystal structure searches using the adaptive genetic algorithm (AGA) \cite{21} to search for iron-rich Fe-O compounds at pressures greater than 100 GPa. Synchrotron XRD experiments on the Fe-O system under high P-T conditions using laser-heated diamond-anvil cells (LHDAC) were carried out at the Advanced Photon Source and the Shanghai Synchrotron Radiation Facility. Starting materials were mixtures of pure iron with FeO or Fe$_2$O$_3$ powder in a size range of submicron to a few microns. We employed repetitive laser pulses of micron seconds for a short period of $\sim$ 1 sec. This laser heating strategy significantly reduced potential carbon diffusion from diamond anvils into the sample chamber \cite{22,23}.

A set of five high P-T experiments were systematically performed to try synthesizing Fe$_n$O compounds. In the first three runs, Fe$_n$O compounds were not formed at 100-200 GPa up to 2,600 K, consistent with the previous experimental studies on FeO \cite{23,24}. In run 4, the Fe-FeO sample mixture was compressed firstly to 220 GPa at room temperature. The relative peak intensities of Fe and FeO suggest that the loaded sample mixture had a Fe/O ratio between 2:1 and 3:1 (Fig. 1). Upon laser pulse heating to 2,000-3,000 K, diffraction peaks of both $\epsilon$-Fe and B8-FeO merely became sharper without the appearance of any new reflections. The sample was heated up readily and evenly on both sides, indicating that it was well insulated from the diamond anvils by the KCl pressure-transmitting medium. Further laser pulse heating to 3,000-3,200 K showed dramatic XRD pattern changes in a few seconds (Fig. 1). The integrated diffraction peak intensities of corresponding Fe and FeO phases rapidly reduced, indicating that iron reacted with FeO at P-T conditions greater than 3,000 K and 220-230 GPa. A similar rapid chemical reaction happened in the run 5 starting from a Fe-Fe$_2$O$_3$ mixture directly compressed to 250 GPa at room temperature and then exposed to repetitive laser pulse heating (Fig. 1). After reaching $\sim$ 3,300 K, the peak intensities of the Fe and Fe$_2$O$_3$ phases decreased, and a set of additional diffraction peaks emerged. The new peaks could not be indexed to high-pressure phases of FeO, Fe$_2$O$_3$, or KCl. At least until 260 GPa and 3,500 K, they remained unaltered before the anvils shattered upon further heating.

\onecolumngrid

\begin{figure}[b]
\includegraphics[width=0.8\textwidth]{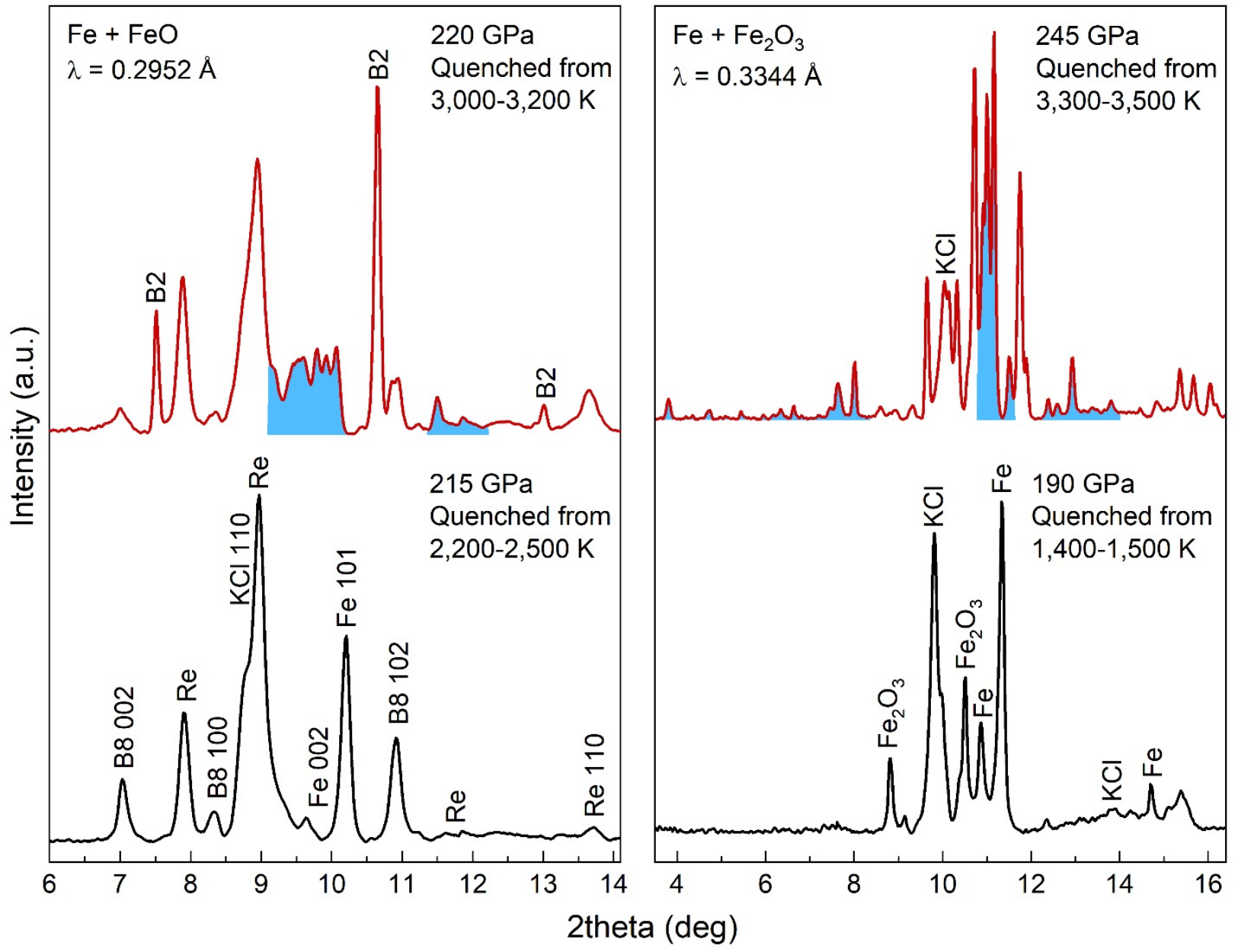}
\caption{\label{fig:fig1} Representative x-ray diffraction patterns collected prior to (bottom) and after (top) the reaction of Fe with FeO at 220 GPa (left) and Fe$_2$O$_3$ at 245 GPa (right) at high temperature. The blue areas highlight some of the new diffraction peaks that emerged at the expense of Fe and FeO/Fe$_2$O$_3$ peaks when the temperature reaching 3,000-3,200 and 3,300-3,500 K at 220 and 245 GPa, respectively. Left panel: B8-FeO has lattice parameters a = 2.3480(17) $\text{\AA}$ and c = 4.7901(41) $\text{\AA}$; hcp-Fe has a = 2.1792(19) $\text{\AA}$ and c = 3.4865(32) $\text{\AA}$; hcp-Re has a = 2.4733 (23) $\text{\AA}$ and c = 3.9944(47) $\text{\AA}$; cubic B2-KCl has a = 2.7292 (53) $\text{\AA}$; B2 denotes B2-FeO. Right panel: starting materials were the mixture of iron and hematite.}
\end{figure}

\clearpage

\begin{figure}[t]
\includegraphics[width=0.88\textwidth]{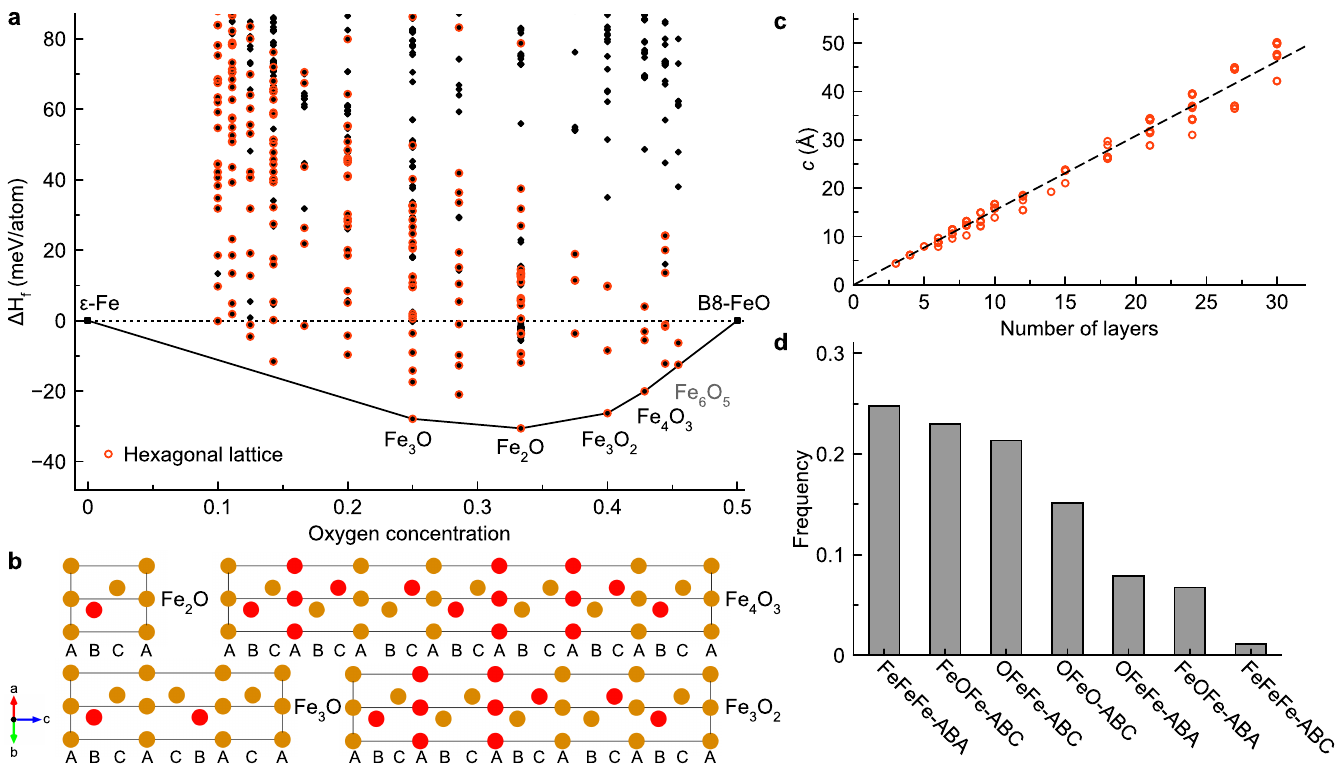}
\caption{\label{fig:fig2}  Crystal structure searches of Fe$_n$O compounds at 215 GPa. (a) The formation enthalpy of AGA-searched compounds referenced by the dashed line shows their relative stability with respect to the decomposition into $\epsilon$-Fe and B8-FeO. The solid line indicates the convex hull formed by the thermodynamically stable compounds. The red symbol highlights structures with a hexagonal lattice. (b) The crystal structures of four Fe-rich compounds in the ground states. Gold circles represent iron while red circles are for oxygen. (c) Lattice parameter $c$ of low-energy close-packed crystals as a function of the number of layers. (d) Frequency of structural and chemical order in nearest neighbor layers of the low-energy close-packed crystals.}
\end{figure}

\twocolumngrid

Crystal structures of Fe-rich Fe$_n$O were searched using the adaptive genetic algorithm (AGA) \cite{21,25,26}, which combines auxiliary interatomic potentials described by the embedded-atom method and ab initio calculations together in an adaptive manner to ensure high efficiency and accuracy. The structure searches were only constrained by the chemical composition, without any assumption on the Bravais lattice type, symmetry, atom basis, or unit cell dimensions. A wide range of different Fe-rich compositions (i.e., 2:1, 3:1, 3:2, 4:1, 4:3, 5:1, 5:2, 5:3, 5:4, 6:1, 6:5, 7:1, 8:1, 9:1) were selected with up to 25 atoms in the unit cell to perform the search. \textit{Ab initio} calculations were carried out using the projector augmented wave (PAW) method \cite{27} within density functional theory as implemented in the VASP code \cite{28,29}. The exchange and correlation energy are treated with the generalized gradient approximation parameterized by the Perdew-Burke-Ernzerhof formula \cite{30}. A plane-wave basis set was used with a kinetic energy cutoff of 650 eV. During the AGA search, the Monkhorst-Pack's sampling scheme \cite{31} was adopted for Brillouin zone sampling with a $\textbf{k}$-point grid of 2$\pi$ $\times$ 0.033 $\text{\AA}^{-1}$, and the ionic relaxations stopped when the forces on every atom became smaller than 0.01 eV/$\text{\AA}$. The energy convergence criterion is 10$^{-5}$ eV. Phonon calculations were performed using density functional perturbation theory \cite{32} implemented in the VASP code and the Phonopy software \cite{33}. The Mermin functional \cite{34,35} is used in all calculations to address the effect of thermal electronic excitation in the metallic states.

Crystal structure searches with AGA at 215 GPa reveal low-energy Fe$_n$O structures between $\epsilon$-Fe and FeO. Fe$_2$O and Fe$_3$O are the two stoichiometric Fe$_n$O compounds with the lowest oxygen concentrations in the convex hull (Fig. 2). A series of Fe$_n$O compounds occurs in the convex hull structures, but their presence can only partially match the XRD peaks of the reaction products. These suggest the presence of other more complex structures in the experimental reaction products. To resolve the XRD patterns with new diffraction peaks at 220 and 245 GPa (Fig. 1), we used large supercells and found that hexagonal disordered structures are lacking long-range order but displaying medium-range structural order at the scale of $\sim$ 30 $\text{\AA}$ with averaged compositions $\sim$  Fe$_2$O (Fe$_{25}$O$_{13}$ and Fe$_{28}$O$_{14}$) that can fit all the additional diffraction peaks (Fig. 3). These results clearly demonstrate the formation of a large family of new Fe$_n$O stoichiometries with related structures in the Fe-FeO system at pressures greater than 200 GPa relevant to Earth's core conditions.

Fig. 2a summarizes the AGA crystal structure searches at 215 GPa. They revealed a large number of new stoichiometric phases more stable than the $\epsilon$-Fe and B8-FeO phases combined. Four compounds with Fe$_3$O (space group $P6_3/mmc$), Fe$_2$O ($P\bar{3}m1$), Fe$_3$O$_2$ ($R\bar{3}m$), and Fe$_4$O$_3$ ($R\bar{3}m$) stoichiometries are ground states, which define the convex hull at this pressure. Fe$_6$O$_5$ ($R\bar{3}m$) is very close to the convex hull with only 0.2 meV/atom above it. Phonon calculations confirm these structures are dynamically stable. Fig. 2b shows that all these new iron-rich compounds have hexagonal close-packed structures, like the end phases $\epsilon$-Fe ($P6_3/mmc$) and B8-FeO ($P6_3/mmc$). Furthermore, most low-energy compounds also have hexagonal close-packed structures, as shown in Fig. 2a (red symbols).

To understand the close-packing motifs in these crystals, we select low-energy structures whose formation enthalpies are within 26 meV/atom ($\sim$ 300 K) above the convex hull. The lattice parameter $c$ of these structures displays an almost linear dependence on the number of layers in the primitive cell, as shown in Fig. 2c, confirming these structures follow a similar layered motif. The nature of these phases can be rationalized by examining the chemical ordering and stacking sequence of the hexagonal close-packed layers. We find only seven different three-layer stacking sequences out of twelve possible ones in these low-energy structures. Fig. 2d shows the frequency of these stackings. The Fe-Fe-Fe layer structure favors the ABA stacking sequence, consistent with the hcp structure in $\epsilon$-Fe. It is much more popular than the ABC stacking sequence counterpart. On the other hand,  layer sequences involving oxygen, e.g., Fe-O-Fe or Fe-Fe-O, favor the ABC stacking over the ABA. Five arrangements involving adjacent O-O layers, e.g., Fe-O-O or O-Fe-O with the ABA stacking sequence, are absent among low-energy structures. Therefore, O layers are well separated from each other. The current structure searches reveal a Fe$_3$O structure with $P6_3/mmc$ symmetry different from previously proposed ones \cite{9,15,16}. The BiI$_3$–type ($R\bar{3}$) structure \cite{15}, once suggested, is also hexagonal, but the oxygen layers are not close-packed. This atomic arrangement violates the close-packing rules found in Fig. 2d and results in a high enthalpy model. Recent computational search for high-pressure iron oxides obtained a Fe3O hexagonal structure ($P\bar{6}m2$) \cite{16} more similar to our $P6_3/mmc$ phase. However, the enthalpy of the $P\bar{6}m2$ phase is $\sim$ 10 meV/atom higher than the $P6_3/mmc$ one. Further calculations up to inner core pressures of the $P6_3/mmc$ Fe$_3$O ground state confirm its stability against phase separation into Fe and FeO.

Comparing the experimental XRD patterns with those of the four ground state phases, we see that none of these phases can reproduce all diffraction peaks of the reaction products. However, a combination of stable and metastable phases provides a better match for the current XRD patterns at 220 GPa and 245 GPa. To clarify this point, we set up Reverse Monte Carlo (RMC) simulations of supercell structures of up to $\sim$ 10 nm in length (60 layers) to search for the Fe-O stacking sequence that best matches the experimental XRD patterns. The interlayer distances in these supercells depend mainly on the chemical ordering of neighboring layers and are modeled using the most likely spacing distances found from the AGA-searched crystal structures. Fig. 3a shows the middle layer spacing distributions for various Fe/O layer sequences.

\begin{figure}[t]
\includegraphics[width=0.48\textwidth]{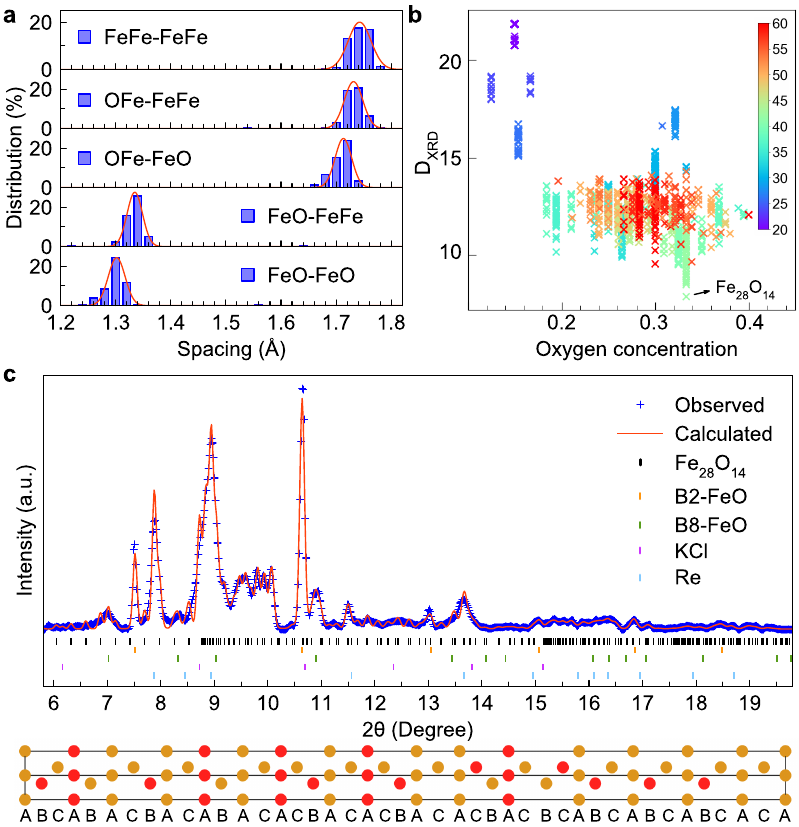}
\caption{\label{fig:fig3}  Reverse Monte Carlo simulation of Fe$_{28}$O$_{14}$ superlattice. (a) Distribution of middle layer spacing with different neighbor layers. The legend indicates the chemical elements in the first and second nearest neighbor layers, e.g., ``OFe-FeO'' denoting that two Fe layers are the first nearest neighbor while two O are the second nearest neighbors. The red curve is a fitting of Gaussian distribution. (b) The oxygen concentration and XRD deviation ($D_{XRD}$) of hexagonal superlattices from RMC simulation. Each point represents the final structure from one RMC simulation. The color bar indicates the number of atoms in the supercell. (c) The comparison of the diffraction pattern between XRD experiments at 220 GPa and Fe$_{28}$O$_{14}$ superlattice from RMC simulation. The incident x-ray wavelength ($\lambda$) is 0.2952 $\text{\AA}$. Vertical ticks: hexagonal Fe$_{28}$O$_{14}$ (black), B2-FeO (orange), B8-FeO (green), B2-KCl (magenta), and hcp-Re (blue). The lower panel shows the simulated crystal structure of Fe$_{28}$O$_{14}$.}
\end{figure}

The XRD deviation, $D_{XRD}$, the criterion used to select preferred structures during the RMC sampling is defined as $D_{XRD}=\sqrt{<(I_{exp}(2\theta)-I_{sim}(2\theta))^2>}$, i.e., the mean square intensity deviation between the simulated, $I_{sim}$, and the experimental, $I_{exp}$, peak intensities. The $2\theta$ range used is 9.2-10.2$^\circ$ at 220 GPa and 10.6-11.6$^\circ$ at 245 GPa. These XRD patterns shown in Figs 1 and 3c include the same significant peaks. Fig. 3b show simulated $D_{XRD}$ for different supercell sizes vs. iron concentrations. The smallest $D_{XRD}$ at 220 GPa is obtained for the 42-layer supercell with chemical composition Fe$_{28}$O$_{14}$ shown in Fig. 3c. This supercell's XRD matches all additional experimental peaks. \textit{Ab initio} calculations confirm this phase is only 28 meV/atom (corresponding to $\sim$ 300 K) above the convex hull at 0 K. The calculated equation of state of this supercell agrees well with the experimental one, confirming that this structure and composition are a reasonable solution for the experimental XRD. A similar RMC calculation was performed to solve the XRD of the reaction products of Fe and Fe$_2$O$_3$ at 245 GPa. The 38-layer supercell solution with composition Fe$_{25}$O$_{13}$ can match the experimental data well. Structural analyses of the supercell solutions indicate they represent hexagonal structures lacking long-range stacking or chemical order. However, they display medium-range order (MRO), i.e., domains with preferred stacking order at the scale of $\sim$ 20 atomic layers or $\sim$ 30 $\text{\AA}$. The combination and interferences between the MRO domains give rise to complex multi-peak patterns in the experimental XRD.

Fig. 4 shows the electronic density of states (DOS) in $\epsilon$-Fe, Fe$_3$O, Fe$_2$O, and B8-FeO at 215 GPa and T$_{\text{el}}$ = 3000 K. All the Fe$_n$O  phases are metallic. In $\epsilon$-Fe, the electronic density of states (DOS) shows a valley at the Fermi level (Fig. 4a). The carrier density increases systematically with increasing oxygen concentration in Fe$_n$O. For instance, oxygen adds carrier at the Fermi level, consistent with being a metallic element at these pressures. Fe$_3$O, for example, has Fe3 and Fe6 type of Fe more distant from oxygen layers and their contribution to the DOS (red area) is similar to $\epsilon$-Fe. In contrast, Fe layers neighboring O monolayers (Fe1, Fe2, Fe4, and Fe5) contribute relatively more carriers to the Fermi level. This trend is maximized in FeO, where all iron layers neighbor oxygen layers, and the Fermi level is at the peak of the DOS and the Fe partial-DOS. Therefore, the presence of oxygen layers affects the electronic DOS and thus electrical and thermal conductivity, essential parameters for understanding the thermal evolution of the core and geodynamo generation in the liquid outer core \cite{36}.

\begin{figure}[t]
\includegraphics[width=0.49\textwidth]{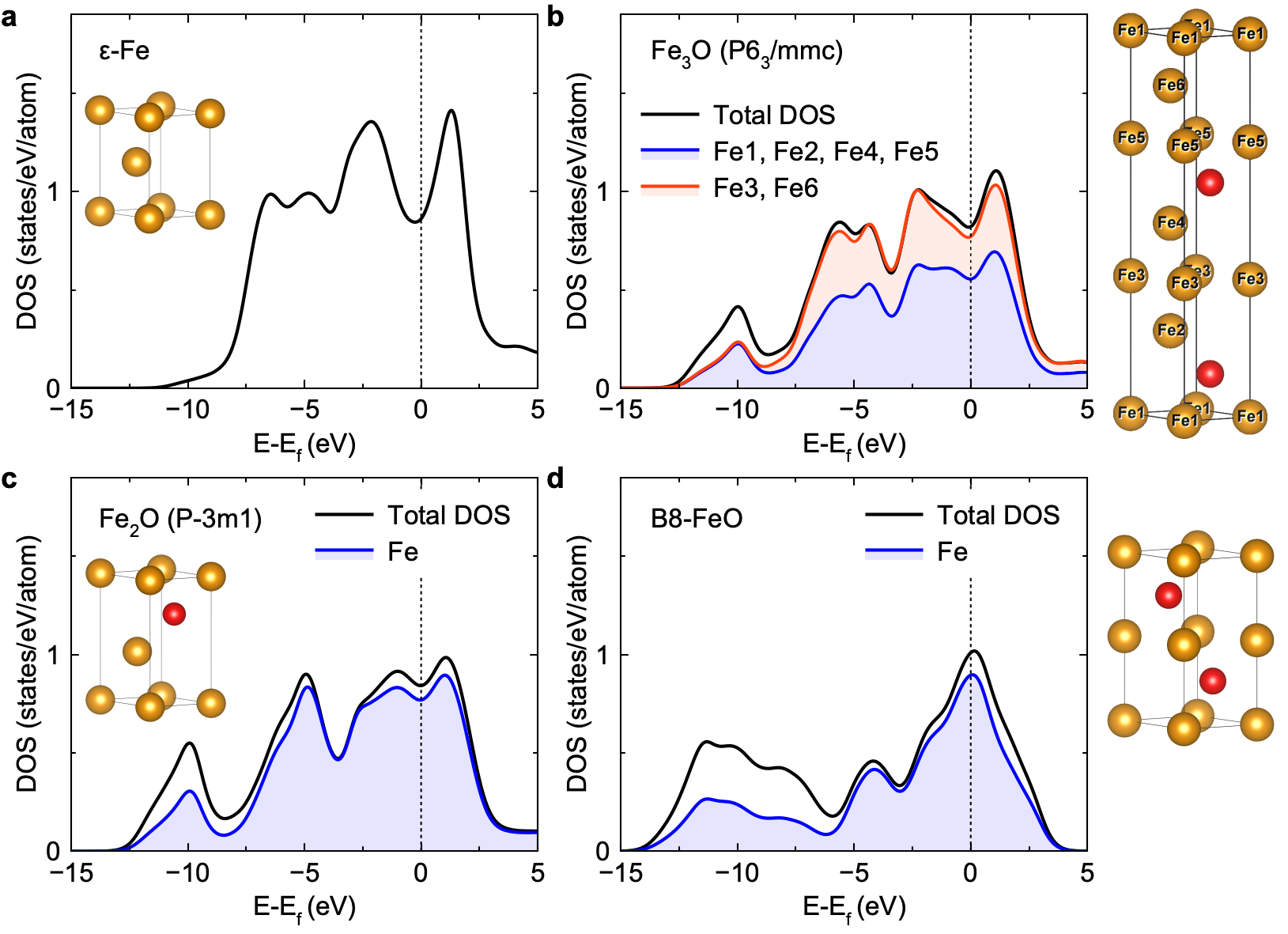}
\caption{\label{fig:fig4}  Projected DOS of Fe $3d$ electrons in Fe and Fe$_n$O compounds at 220 GPa and T$_{\text{el}}$ = 3000K. (a) $\epsilon$-Fe. The dashed line indicates the Fermi level. (b) Fe$_3$O (space group $P6_3/mmc$). The red area shows the partial DOS of Fe atoms (Fe3 and Fe6) with no oxygen first neighbors. Their contribution to the DOS is similar to $\epsilon$-Fe. The blue area shows the relatively larger contribution of Fe with oxygen first neighbors (i.e., Fe1, Fe2, Fe4, and Fe5) to the carrier density. (c) Fe$_2$O ($P\bar{3}m1$). (d) B8-FeO. Gold and red spheres in the crystal structure represent iron and oxygen, respectively. }
\end{figure}

In summary, high-pressure experiments and crystal structure searches combined reveal that the hexagonal close-packed $\epsilon$-Fe structure can incorporate oxygen at 220-260 GPa and temperatures greater than 3000 K. The Fe$_n$O  phases identified in this work are greatly stabilized by the formation of close-packed oxygen-only monolayers with respect to metallic iron and oxygen at megabar pressures. This type of alloying is consistent with the B8 structure of metallic FeO at high pressures, which has alternating close-packed layers of oxygen and iron. Compared to the previously proposed random configuration of oxygen defect structures in solid iron  \cite{9,15}, the oxygen-only monolayer model presented here can significantly minimize lattice strain and over-coordination energetic cost. The current Fe$_n$O phases are also different from the Fe$_n$S phases Fe$_3$S and Fe$_3$S$_2$ that can have Fe and S, another column VI element, co-existing in the same layer \cite{8, 9, 37}. Since oxygen is believed to be a likely light element alloyed with Fe in the Earth's liquid outer core \cite{38,39,40}, the synthesis of solid iron-rich Fe$_n$O phases suggests that oxygen could also be a likely light element in the inner core. Future studies on how oxygen interacts with other light elements in dissolved iron and oxygen partitioning between the solid inner and liquid outer core will help clarify the origin of the density deficit and the light elements present in the Earth's core.

\begin{acknowledgments}
This work is supported by the NSFC Grants no. 42072052 and U1930401, by National Science Foundation awards EAR-1918126. (R.M.W. and Y.S.), EAR-1918134 (K.-M.H. and C.Z.W), and EAR-1901808 and EAR-1916941 (J.-F.L.). R.M.W. also acknowledges partial support from the Department of Energy, Theoretical Chemistry Program through grant DE-SC0019759. Computational resources were provided by the Extreme Science and Engineering Discovery Environment (XSEDE) funded by the National Science Foundation through award ACI-1548562. This research also used resources of the Advanced Photon Source (A.P.S.), a U.S. Department of Energy (D.O.E.) Office of Science User Facility operated by Argonne National Laboratory under Contract No. DE-AC02-06CH11357. The GeoSoilEnviroCARS at A.P.S. is supported by the National Science Foundation - Earth Sciences award EAR-1634415 and the Department of Energy-GeoSciences award DE-FG02-94ER14466.
\end{acknowledgments}

\bibliographystyle{apsrev4-2}


\end{document}